\documentclass[12pt,reqno]{amsart}
\usepackage{amsmath,amssymb,amsfonts,amsthm}
\usepackage[mathscr]{eucal}
\usepackage[all]{xy}
\label{}
\usepackage[english]{babel}
\usepackage{cite}

\usepackage{hyperref}
\begin{document}

%-------------------------------------------------------------------------
% editorial commands: to be inserted by the editorial office
%
%\firstpage{1} \volume{228} \Copyrightyear{2004} \DOI{003-0001}
%\seriesextraline{This is the Concrete Title of this Book\br H.E. R and S.T.C. W, Eds.}
%\firstpage{1}

\title[Energy and stability of Pais-Uhlenbeck oscillator]
 {Energy and stability of Pais-Uhlenbeck oscillator}

\author{D.S. Kaparulin and  S.L.  Lyakhovich}

\address{Department of Quantum Field Theory,
Tomsk State University, Lenin ave. 36, Tomsk 634050, Russia.}

\email{dsc@phys.tsu.ru, sll@phys.tsu.ru}

\thanks{The work is partially supported by the Tomsk State
University Competitiveness Improvement Program and the RFBR grant
13-02-00551-a.  S.L.L. is partially supported by the RFBR grant
14-01-00489-a, D.S.K. is grateful for the support from Dynasty
Foundation}

\begin{abstract} We study stability of higher-derivative dynamics
from the viewpoint of more general correspondence between symmetries
and conservation laws established by the Lagrange anchor. We show
that classical and quantum stability may be provided if a
higher-derivative model admits a bounded from below integral of
motion and the Lagrange anchor that relates this integral to the
time translation.

\end{abstract}

\maketitle

\section*{Introduction}

A notorious trouble appears when the Noether theorem \cite{K-S} is
applied to the theories with higher derivatives, the models whose
Lagrangians depend on accelerations and higher derivatives of
generalized coordinates. In contrast to the-lower order theories,
where unboundedness of the canonical energy from below usually
indicates the presence of ghost states and instability of the model,
in higher-derivative theories the unboundedness of canonical energy
is not necessary to have negative impact on classical dynamics
\cite{Smilga2005,Smilga2014}. The relationship between
(un)boundedness of canonical energy from below and (in)stability of
higher-derivative theory was subject of many works
\cite{Chen2013,BENDER2007,BOLONEK2005,DAMASKINSKY2006,Mostafazadeh2010}.

In this note, we consider a stability of higher-derivative dynamics
from the viewpoint of more general correspondence between symmetries
and conservation laws which is established by the Lagrange anchor.
Following the ideas of \cite{KapLS2014EPhysC,KapLS2014RussPhys}, we
show that the stability of higher-derivative theory may be provided
if the model admits a bounded from below integral of motion and a
Lagrange anchor that associates the integral of motion with
translation in time. The general construction is illustrated by the
example of the Pais-Uhlenbeck oscillator.

The paper is organized as follows. In Section \ref{1}, we recall
some basic facts about the conservation laws of the Pais-Uhlenbeck
(PU) oscillator. In Section \ref{2}, we introduce the Lagrange
anchor for the PU oscillator and establish a correspondence between
symmetries and conservation laws. The bounded integral of motion and
the Lagrange anchor that associates it to time translations are
identified. Section \ref{3} is devoted to the Pais-Uhlenbeck
oscillator with equal frequencies. We show that in unstable theory
the bounded integral of motion exists, but it appears to be
unrelated to the time-translation symmetry.

\section{Conservation laws of the PU oscillator}\label{1}

We consider the one-dimensional Pais-Uhlenbeck oscillator of order
$2n$ \cite{PU}, whose action functional has the form
\begin{equation}\label{PUL}
S[x(t)]= \int dt L\,, \qquad
L=\frac{1}{2\Omega}x(t)\prod_{i=1}^n\Big(\frac{d^2}{dt^2}+\omega_i^2\Big)x(t)\,;
\end{equation}
here
$$0<\omega_1<\omega_2<\ldots<\omega_n$$
are the frequencies of oscillations. We assume that there is no
resonance, so that all the frequencies are different. We also
introduced the dimensional constant $\Omega>0$ to provide the
correct dimension of the action (\ref{PUL}).

The corresponding equation of motion reads
\begin{equation}\label{LEPU}
    \frac{\delta S}{\delta x}\equiv\frac{1}{\Omega}
    \prod_{i=1}^n\Big(\frac{d^2}{dt^2}+\omega_i^2\Big)x=0\, .
\end{equation}
It is convenient to introduce the wave operator that defines r.h.s.
of the equation of motion
\begin{equation}\label{Top}
T=\frac{1}{\Omega}
    \prod_{i=1}^n\Big(\frac{d^2}{dt^2}+\omega_i^2\Big)\,,\qquad
    \frac{\delta S}{\delta x}=T(x)\,.
\end{equation}
We will also use notation
$$
\dot{x}=\frac{dx}{dt},\qquad\ddot{x}=\frac{d^2x}{dt^2},\qquad\ldots,\qquad\stackrel{_{(n)}}{x}=\frac{d^nx}{dt^n}\,
$$
for the time derivatives of $x$.

The general solution of equation~(\ref{LEPU}) is given by the sum of
$n$ oscillations with frequencies $\omega_i$ with different
amplitudes $A_i$ and phases $\varphi_i$
\begin{equation}\label{sumxi}
x(t)=\sum_{i=1}^n x_i(t)\,,\qquad x_i(t)\equiv A_i\sin(\omega_i
t+\varphi_i)=\mathcal{P}_i x(t)\,.
\end{equation}
Here, the operators
$$
\mathcal{P}_i=\prod_{j\neq i}\frac{1}{\omega_j^2-\omega_i^2}
\Big(\frac{d^2}{dt^2}+\omega_j^2\Big)
$$
have the sense of projectors to the subspaces of solutions with
frequencies $\omega_i$, respectively. There are two obvious
properties
\begin{equation}\label{sumP}\begin{array}{c}\displaystyle
\sum_{i=1}^n\mathcal{P}_i=1\,,\\[3mm]
\displaystyle\mathcal{P}_i^2=\mathcal{P}_i-\Omega\sum_{j\neq
i}\prod_{r\neq i}\frac{1}{\omega_i^2-\omega_r^2}\prod_{s\neq
j}\frac{1}{\omega_j^2-\omega_s^2}\prod_{k\neq
j,i}\Big(\frac{d^2}{dt^2}+\omega_k^2\Big)T\,.
\end{array}\end{equation}
In order to prove the first relation one can apply to it the Fourier
transform:
$$
F(\sum_{i=1}^n\mathcal{P}_i-1)=\sum_{i=1}^n \prod_{j\neq
i}\frac{1}{\omega_j^2-\omega_i^2} \Big(\omega_j^2-p^2\Big)-1\,.
$$
Then the r.h.s. of the relation is a polynomial of degree $2n-2$
that has $2n$ roots $p=\pm\omega_i$ and thus has to be equal to zero
identically. The second relation follows from identity
$$
\mathcal{P}^2_i=\mathcal{P}_i\Big(1-\sum_{j\neq i}\mathcal{P}_j\Big)
$$
with account of notation (\ref{Top}).

Due to relations (\ref{sumP}), formula (\ref{sumxi}) establishes a
correspondence between the solutions to the PU oscillator equation
and the system of $n$ independent harmonic oscillators
\begin{equation}\label{x-xi}
\frac{\delta S}{\delta x(t)}=0\qquad\Leftrightarrow\qquad
\Big(\frac{d^2}{dt^2}+\omega_i^2\Big)x_i(t)=0,\qquad i=1,2,\ldots,n.
\end{equation}
From (\ref{x-xi}) it immediately follows that the PU oscillator has
$n$ independent integrals of motion
\begin{equation}\label{Ii}
I_i\equiv\frac{1}{2}\Big[\dot{x}_i^2+\omega_i^2x_i^2\Big]=
\frac{1}{2}\Big[\Big(\mathcal{P}_i
\dot{x}\Big)^2+\omega_i^2\Big(\mathcal{P}_i x\Big)^2\Big]\,.
\end{equation} %If the ratio of some frequencies is rational
%$\omega_i/\omega_j=m/n$ for some $i\neq j$, the PU oscillator has
%additional integrals of motion \cite{}, but this is not important
%for our consideration.

The general quadratic integral of motion is given by the linear
combination of integrals (\ref{Ii}). Namely,
\begin{equation}\label{sumIi} I=-\sum_{i=1}^n
\Big[\prod_{j\neq
i}\Big(\omega_j^2-\omega_i^2\Big)\Big]\frac{\alpha_i I_i}{\Omega}\,,
\end{equation}
with $\alpha_i$ being arbitrary real constants.

It is easy to see that expression $I$ in~(\ref{sumIi}) is conserved
\begin{equation}\label{dI}
\frac{dI}{dt}=Q\frac{\delta S}{\delta x}\,,\qquad
Q(t,x,\dot{x},\ldots,\stackrel{_{(2n-1)}}{x})=
-\sum_{i=1}^n\Big(\alpha_i\mathcal{P}_i\dot{x}\Big)\,,
\end{equation}
where the coefficient $Q$ is called the \emph{characteristic}
associated with the conservation law $I$. It is known
\cite{KapLS2010} that there is a one-to-one correspondence between
integrals of motion and characteristics. The last fact allows one to
use characteristics for establishing a correspondence between the
symmetries and conservation laws. The simplest example is provided
by the Noether theorem. The Noether theorem identifies the
characteristic $Q$ with the infinitesimal symmetry transformation of
action functional:
\begin{equation}\label{Qtransf}
    \delta_\varepsilon x = \varepsilon Q \, , \quad \delta_\varepsilon S=0\qquad
    \Leftrightarrow\qquad
    \frac{dI}{dt}=Q\frac{\delta S}{\delta x}\,.
\end{equation}

The problem appears when a conservation law bounded form below is
associated with the time translations. The general bounded
conservation law~(\ref{sumIi}) with $(-1)^i\alpha_i>0$ corresponds
to some symmetry of the action (\ref{PUL})
\begin{equation}\label{dx}
    \delta_\varepsilon x=-\varepsilon
    \sum_{i=1}^n\Big(\alpha_i\mathcal{P}_i\dot{x}\Big)\,,
\end{equation}
while the infinitesimal time translation $\delta_\varepsilon
x=-\varepsilon \dot{x}\,$ corresponds to the unbounded conservation
law with $\alpha_i=1$\,. This is manifestation of general no-go
statement about unboundedness of energy in the theories with higher
derivatives. Unless the higher-derivative theory is highly
constrained, the usual Noether theorem can't connect a positive
conserved quantity to the time translation invariance, see for
instance discussion in \cite{Chen2013} and references therein.

\section{The Lagrange anchor for the PU oscillator}\label{2}

The generalization of the Noether theorem suggests that the
correspondence between the symmetries and characteristics (and hence
conservation laws) is established by the linear differential
operator\footnote{The notation of this section is adapted for the
case of the PU oscillator. For general definitions of the Lagrange
anchor and correspondence between symmetries and conservation laws
see \cite{KapLS2010,KazLS}.}
$$
V=\sum_{i=1}^{2n-1}\stackrel{_{(i)}}{V}(t,x,\dot{x},\ldots,\stackrel{_{(2n-1)}}{x})
\frac{d^i}{dt^i}\,
$$
that associates the characteristic $Q$ with the symmetry
\begin{equation}\label{GNT}
\delta_\varepsilon x=\varepsilon
V(Q)=\sum_{i=1}^{2n-1}\stackrel{_{(i)}}{V}\frac{d^iQ}{dt^i}\,,\qquad\delta_\varepsilon\Big(\frac{\delta
S}{\delta x}\Big)\Big|_{\frac{\delta S}{\delta x}=0}=0\,.
\end{equation}
Invariance of the equation of motion under transformation~(\ref{GNT}) implies certain compatibility condition for the Lagrange
anchor \cite{KapLS2010}. In the simplest case of linear equations
$\delta S/\delta x=T(x)$ and the Lagrange anchor with no field and
time dependence,
$$
\stackrel{_{(i)}}{V}=\mathrm{const},\qquad i=0,\ldots,2n-1\,,
$$ this compatibility condition
takes the form \cite{KapLS2014EPhysC}
\begin{equation}\label{LA}
V^*T^*-TV=0\,.
\end{equation}
For a self-adjoint wave operator $T=T^\ast$ (which is always true
for Lagrangian theories) and a self-adjoint Lagrange anchor
$V=V^\ast$, relation (\ref{LA}) takes even more simple form
\begin{equation}\label{VT}
[V,T]=0\,.
\end{equation}
The obvious solution $V=1$ corresponds to the canonical Lagrange
anchor which is always admissible for Lagrangian theory. It
establishes the Noether correspondence between symmetries and
conservation laws (\ref{Qtransf}). The canonical Lagrange anchor
can't connect bounded from below integral of motion with translation
in time, we have to be interested in the non-canonical Lagrange
anchors. We have the following $n$-parameter family of the Lagrange
anchors for the PU oscillator of order $2n$:
\begin{equation}\label{VPU}
V=\sum_{i=1}^n\beta_i\mathcal{P}_i\,,
\end{equation}
with $\beta_i$ being arbitrary real constants. The details about
derivation of this Lagrange anchor can be found in
\cite{KapLS2014EPhysC}.

The Lagrange anchor (\ref{VPU}) associates the general conservation
law (\ref{sumIi}) with the symmetry
\begin{equation}\label{VQ} \delta_\varepsilon
x=\varepsilon
V(Q)=-\varepsilon\sum_{i=1}^n\Big(\alpha_i\beta_i\mathcal{P}_i\dot{x}\Big)+
\varepsilon\Omega R(\alpha_i,\beta_j)T(x)\,,
\end{equation}
where
$$
R(\alpha_i,\beta_j)=\sum_{i,j=1}^n\Big(\alpha_i\beta_i-\alpha_i\beta_j\Big)\prod_{r\neq
i}\frac{1}{\omega_i^2-\omega_r^2}\prod_{s\neq
j}\frac{1}{\omega_j^2-\omega_s^2}\prod_{k\neq
j,i}\Big(\frac{d^2}{dt^2}+\omega_k^2\Big)\,.
$$
The second term in (\ref{VQ}) is given by the linear combination of
equations of motion and their differential consequences, and thus
should be considered as trivial. Below, we will consider symmetries
modulo trivial terms.

The crucial difference between Noether's correspondence
(\ref{Qtransf}) and (\ref{VQ}) is that the symmetry (\ref{VQ})
depends on $n$ free parameters. We can use this ambiguity of the
Lagrange anchor to connect the general integral of motion
(\ref{sumIi}) with translations in time. Whenever $\alpha_i\neq0$
the desired correspondence
\begin{equation}\label{dotx}
V(Q)=-\dot{x}+\Omega R(\alpha_i,1/\alpha_j)T(x)\,
\end{equation}
is established for
\begin{equation}\label{beta=}
\beta_i=\frac{1}{\alpha_i}\,.
\end{equation}

In contrast to the Noether energy, the conservation law
(\ref{sumIi}) can be bounded or unbounded from below depending on
the value of $\alpha$'s. The bounded integrals of motion
(\ref{sumIi}) are associated with time translations by differential
Lagrange anchor
\begin{equation}\label{LA+}
    V=\sum_{i=1}^n\frac{1}{\alpha_i}\mathcal{P}_i,\qquad
    (-1)^i\alpha_i>0\,.
\end{equation}
From the classical viewpoint the relationship (\ref{GNT}) is as good
as the Noether's one. In particular, it allows one to define the
generalization of the Dickey bracket of conservation laws and admits
BRST-description \cite{KapLS2011JHEP}. Thus, the correspondence
between the bounded from below conservation law (\ref{sumIi}), the
Lagrange anchor (\ref{LA+}) and the time translation (\ref{dotx})
ensures the stability of the PU oscillator theory even if the
canonical energy is unbounded.

Let us give one more argument that makes analogy between the energy
and conservation law associated with the time translation more
explicit. It is well known that different Lagrange anchors result in
different quantizations of one and the same classical system
\cite{KazLS, LS1}. In the first-order formalism, the
integrable\footnote{ See the definition of integrability in
\cite{KapLS2010}. The field-independent Lagrange anchor (\ref{VPU})
is automatically integrable.} Lagrange anchor always defines a
Poisson bracket on the phase space of the system, while the
corresponding integral of energy becomes the Hamiltonian
\cite{KazLS,KapLS2013,BG2011a}. The canonical Lagrange anchor
corresponds to the canonical Poisson brackets and Hamiltonian that
follows from the Ostrogradsky formalism \cite{Ost,GLT}, while
non-canonical Lagrange anchors correspond to non-canonical Poisson
brackets and Hamiltonians.

\section{The case of resonance}\label{3}

Let us consider the 4-th order PU oscillator in the case of equal
frequencies $\omega=\omega_1=\omega_2$. The equation of motion reads
\begin{equation}\label{PUR}
T(x)=\Big(\frac{1}{\omega}\frac{d^2}{dt^2}+\omega\Big)^2x=0\,,\qquad
T=\Big(\frac{1}{\omega}\frac{d^2}{dt^2}+\omega\Big)^2\,.
\end{equation}
The solutions to the equation of motion demonstrate runaway behavior
with linear time dependence of the oscillation amplitude
$$
x(t)=A\sin(\omega t+\varphi_0)+B t \sin(\omega t+\varphi_1)\,
$$
with $A$, $B$, $\varphi_0$ and $\varphi_1$ being arbitrary real
constants. The system (\ref{PUR}) still has two impendent integrals
of motion
$$
I_1=\frac{1}{2}\Big(\frac{1}{\omega^2}\dddot{x}+\dot{x}\Big)^2+
\frac{1}{2}\Big(\frac{1}{\omega}\ddot{x}+\omega x\Big)^2\,,\quad
I_2=\frac{1}{2}\left(
\frac{\ddot{x}{}^2-2\dddot{x}\dot{x}}{\omega^2}  - 2
    \dot{x}{}^2 -\omega^2x^2 \right)\,.
$$
The first integral is obtained from (\ref{Ii}) by taking limit
$\omega_1\rightarrow\omega_2$ with special renormalization of the
$\alpha-$constants. The second one is just the Noether energy. In
contrast to the case of unequal frequencies, it is impossible to
find two independent bounded from below quadratic integrals of
motion. Only the integral of motion $I_1$ is bounded from below.
However, an attempt to associate it with time translation fails.

The characteristic for $I_1$ reads
$$
\frac{dI_1}{dt}=Q_1T(x),\qquad
Q_1=\Big(\frac{1}{\omega^2}\dddot{x}+\dot{x}\Big)\,.
$$
There are two-parameter family of Lagrange anchors for PU oscillator
(\ref{PUR})
\begin{equation}\label{LAR}
    V=\frac{\beta_2}{\omega^2}\frac{d^2}{dt^2}+\beta_1\,.
\end{equation}
The corresponding symmetry reads
\begin{equation}\label{SymR}
    V(Q_1)=\frac{\beta_1}{\omega^2}T(x)+
    \frac{\beta_2-\beta_1}{\omega^2}\dddot{x}+
    (\beta_2-\beta_1)\dot{x}\,.
\end{equation}
In (\ref{SymR}), the third derivative vanishes if and only if the
symmetry (\ref{SymR}) is trivial, i.e., $\beta_1=\beta_2$. In view
of above, there is no time-independent bounded from below
conservation law that could be associated to time translation. This
result demonstrates the fact that has been already observed in
\cite{Mostafazadeh2010}, where it was found that PU oscillator with
resonance does not admit Hamiltonian formulation with any bounded
Hamiltonian.

\section*{Conclusion}

We observe that for higher-derivative theories, the stability does
not necessarily require the Noether energy to be  bounded from
below. The classical stability can be ensured by a weaker condition
that the model admits a bounded integral of motion.  Once the
equations of motion admit the Lagrange anchor such that maps the
bounded integral to the time translation, the theory can retain
stability at quantum level. Both the conserved quantity and the
Lagrange anchor are not uniquely defined by the equations of motion
and may exist even in non-singular models. This allows us to expand
the stability analysis to the wide class of higher-derivative
theories, including non-singular ones. The general idea is
exemplified by the Pais-Uhlenbeck oscillator. Using the ambiguity of
choice of the Lagrange anchor and bounded conserved quantity, we
demonstrate the stability of Pais-Uhlenbeck oscillator when all the
frequencies are different.

\end{document}